# Guest-Tunable Dielectric Sensing Using a Single Crystal of HKUST-1

Arun S. Babal,[a] Abhijeet K. Chaudhari,[a] Hamish H. -M. Yeung,[b,c] and Jin-Chong Tan[a,*]

[a]Multifunctional Materials and Composites (MMC) Laboratory, Department of Engineering Science, University of Oxford, Parks Road, Oxford, OX1 3PJ, UK

[b]Inorganic Chemistry Laboratory, University of Oxford, South Parks Road, Oxford, OX1 3QR, UK

[c]University of Birmingham, School of Chemistry, Haworth Building, Edgbaston, Birmingham, B15 2TT, UK

*E-mail: jin-chong.tan@eng.ox.ac.uk

**ABSTRACT:** There is rising interest on low-$k$ dielectric materials based on porous metal-organic frameworks (MOFs) for improved electrical insulation in microelectronics. Herein, we demonstrate the concept of MOF dielectric sensor built from a single crystal of HKUST-1. We study guest encapsulation effects of polar and non-polar molecules, by monitoring the transient dielectric response and AC conductivity of the crystal exposed to different vapors (water, $I_2$, methanol, ethanol). The dielectric properties were measured along the <100> crystal direction in the frequency range of 100 Hz to 2 MHz. The dielectric data show the efficacy of MOF dielectric sensor for discriminating the guest analytes. The time-dependent transient response reveals dynamics of the molecular inclusion and exclusion processes in the nanoscale pores. Since dielectric response is ubiquitous to all MOF materials (unlike DC conductivity and fluorescence), our results demonstrate the potential of dielectric MOF sensors compared to resistive sensors and luminescence-based approaches.

Metal-organic frameworks (MOFs) are open-framework compounds that feature interconnected functionally active pores with vast chemical and physical tunability. Because of their chemical structure and highly porous architecture, indeed the majority of MOFs are excellent electrical insulators. Therefore there is a growing interest to study the 'low-$k$' dielectric properties of MOFs,[1-2] as a next-generation dielectric for replacing conventional silicon dioxide (whose real part of dielectric constant, $\varepsilon' = 3.9$-$4.3$), or silicon nitride ($\varepsilon' = 6.0$-$7.0$) prevalent in microelectronics today.[3] Low-$k$ dielectrics are important for boosting the performance of miniaturized integrated circuits (IC), reducing losses due to power dissipation, electronic crosstalk, and resistive-capacitive delays of ICs. Low-$k$ materials can be designed by introducing air pockets into solids, because air (or vacuum) has the lowest attainable $\varepsilon' \sim 1$.[4] Of course, this feature is easily fulfilled by MOF-type materials given their high porosity and low framework density.[5-6] However, outstanding challenges encompass interface adhesion, mechanical, chemical, and thermal stability to afford practical applications.[7]

All the MOF dielectric experiments reported, however, were determined from powder samples (in pressed pellet form) that may suffer from structural collapse and/or framework amorphization from pressure.[8-9] Several studies have employed polycrystalline thin-film samples,[10-11] where grain boundaries defects may be prevalent. Yet, the dielectric property of a single crystal of MOF has not been measured directly, to yield an intrinsic value independent of pelleting pressure and free of grain boundary influence. Importantly, porous MOFs can readily accommodate guest molecules that will alter their dielectric properties by means of host-guest interactions. These physicochemical alterations are molecule specific, and thus can be exploited to achieve highly selective sensors. One way to detect host-guest interactions is by monitoring the transient response of the electrical signals through capacitive impedance or alternating current (AC) conductivity measurements.[12] We envisage that encapsulated polar molecules (as the 'guest') will add an extra dipole to the MOF framework (as the 'host'), thereby enabling highly responsive detection of analytes when sensing subtle changes by leveraging transient dielectric response of a single crystal. Importantly, because all MOFs exhibit a dielectric response, this proposed sensing approach will have major advantages over competitive techniques in the field, including resistive sensing[13-14] or fluorescent sensing[15-16] that require an electrically conductive framework (uncommon for MOFs)[17-18] or luminescence (not true for all MOFs, unlike the dielectric response).

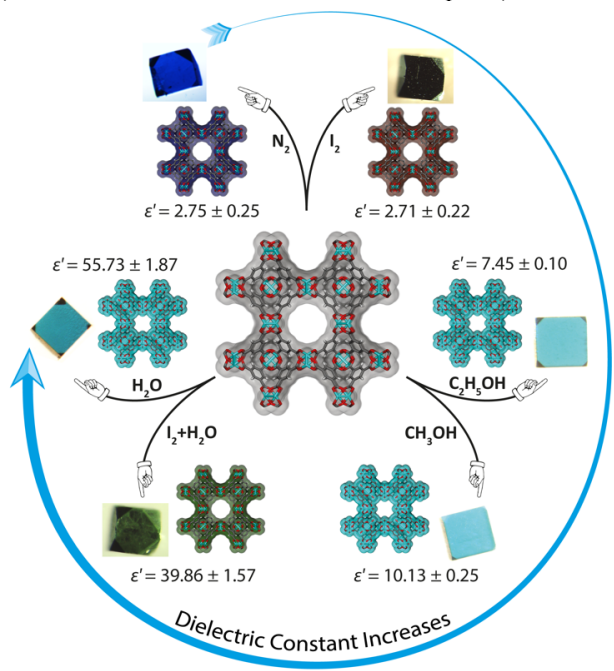

Figure 1: Single-crystal dielectric properties of HKUST-1 at 1 MHz, demonstrating its dielectric ($\varepsilon'$) tuneability through the encapsulation of specific guest molecules. The highest dielectric value was detected for $H_2O$@HKUST-1, while other guest-loaded crystals exhibited a rising trend of $\varepsilon'$ as indicated by the arrow. Photographs show the actual HKUST-1 single crystals tested, whose size were ~250-300 μm.

To probe the effects of guest inclusion on the dielectric properties of a MOF crystal, we have chosen the copper-based MOF, termed HKUST-1 [$Cu_3(BTC)_2$, BTC = benzene-1,3,5-tricarboxylic acid], to serve as a porous 'host' framework because of its ability to form relatively large and mechanically stable single crystals for this study (Figures 1 & 2). Notably, the coordinatively unsaturated metal sites (CUS) in HKUST-1 are ideal adsorption sites for binding various guest molecules.[19-20] We study the effects of both polar (water, methanol and ethanol) and non-polar molecules (iodine, $I_2$) on the dielectric properties of the HKUST-1 crystal, in the frequency range of 100 Hz to 2 MHz at room temperature. We characterized dielectric response of individual crystals along the <100> direction, in its activated state and guest encapsulated states; the primary results are summarized in Figure 1. Subsequently, single-crystal cyclic sensing tests were performed at 1 MHz to probe the transient dielectric response and host-guest dynamics during the sorption/desorption cycles of guest molecules.

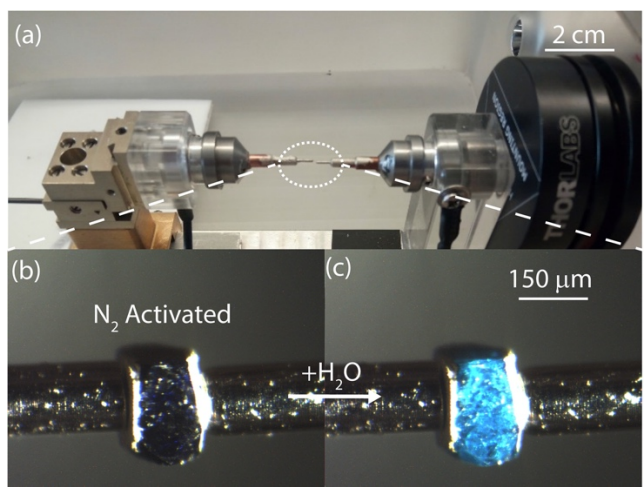

Figure 2: (a) A custom-built single-crystal dielectric probe setup comprising a pair of spring-loaded electrodes mounted on a 3D translational stage. The entire setup is positioned in an enclosed chamber to permit the controlled flow of different guest species (see Figure S1). (b, c) Color change of an activated HKUST-1 single crystal when it was subjected to moisture in air.

Synthesis of the HKUST-1 single crystals was carried out using the solvothermal method described in the SI (§1.1). The cuboidal crystals obtained were characterized under the stereo optical microscope, and found to have of a lateral dimension of ~250-300 μm and a thickness of ~150 μm. A thin layer of silver coating was applied onto the opposite surfaces of the (100)-oriented facets (Figure S2), to achieve a parallel-plate configuration for dielectric measurements (Figure S3). Figure 2 shows the electrode arrangement of the single-crystal dielectric setup. The deep blue color of the pristine crystal (after $N_2$ activation) transforms into light blue on exposure to water vapor, causing change in its refractive index.

Figure 1 presents a summary of the results showing the variation of the guest-dependent optical and dielectric properties of the HKUST-1 crystal, with its dielectric constant tunable between $\varepsilon' =$ 2.8–55.6. The activated crystal switches from deep blue to light blue upon moisture adsorption, accompanied by a ~20-fold rise in its dielectric constant. Non-polar iodine molecules ($I_2$) were encapsulated into the framework by vapor diffusion of sublimated iodine, resulting in color alteration of the HKUST-1 crystal, from deep blue to brown in dehydrated state, and to light green in hydrated state. When exposed to methanol ($CH_3OH$) and ethanol ($C_2H_5OH$) vapors, the crystal switches to a transparent light blue.

Figure S4 shows the X-ray diffraction (XRD) pattern of the HKUST-1 single crystal, where the (200)- and (400)-oriented planes were confirmed. Analysis of the XRD peak profiles (Figure S5) revealed an increase in FWHM of the iodine-encapsulated crystal, indicative of strain developed in framework from adsorbed iodine. Raman spectroscopy was performed to validate the encapsulation of molecular iodine in HKUST-1 (Figure S6). The distinct Raman band at ~210 cm$^{-1}$ can be assigned to the $I_2$ molecules, while the peak at ~420 cm$^{-1}$ is the first overtone of $I_2$.[21] These results confirmed that $I_2$@HKUST-1 crystals were successfully obtained.

Figure 3 shows the influence of different guest molecules on the dielectric properties of HKUST-1 under cyclic sorption and desorption conditions, measured as a function of frequency up to 2 MHz. The pristine crystals were activated in $N_2$ (Figure 3b) before commencing the guest influx experiments. Guest inclusion into the porous framework was carried out by flowing $N_2$ through a bubbler containing the guest species (as liquid) to create a vapor phase. The dielectric constant of the HKUST-1 crystal in its activated state was found to be $\varepsilon' = 2.95$ at 1 MHz frequency (Figure 3a) determined in the <100> lattice direction. However, we observed a minor decline in $\varepsilon'$ with further evacuation and guest inclusion cycles likely due to strain-induced cracking in the crystal interior; inclusion of microscopic air gaps may decrease $\varepsilon'$ (Figure S3). The dielectric constant of the HKUST-1 crystal increases dramatically with $H_2O$ sorption, up to the value of $\varepsilon' = 57$ for 30% RH and $\varepsilon' = 64$ for 70% RH at 1 MHz (see Figure 3a and Figure S10). The increase in $\varepsilon'$ is associated with the inherent dipole of the guest water molecule. Water molecules coordinate to CUS of the copper-paddlewheel in HKUST-1 framework, whereas uncoordinated molecules in the pore will form clusters *via* hydrogen bonding (maximum of 186 molecules per unit cell).[22] We found only a modest rise of $\varepsilon'$ at 70% RH (Figure S10), indicating there is good water occupancy even at 30% RH.

The non-polar molecular iodine ($I_2$) was sublimated into gaseous form and then encapsulated into HKUST-1 to yield the $I_2$@HKUST-1 crystal. At this stage, although HKUST-1 framework has a higher propensity to adsorb $I_2$ over water, the presence of coordinated water on CUS will impede the direct interaction between $I_2$ and the copper centers. Additional preferential interaction sites for molecular iodine are carbonyl oxygen atoms in the copper paddlewheel (3.23−3.30 Å) and the BTC linker (3.47−4.31 Å).[23] Figure 3b shows that in the first water inclusion cycle of $I_2$@HKUST-1, the value of $\varepsilon'$ increased from 2.55 to 37.94 at 30% RH. The maximum value recorded is thus well below the value of $\varepsilon' = 57$ achieved by iodine-free HKUST-1 at 30% RH (Figure 3a). This result reveals that due to the preferential adsorption sites of molecular $I_2$ near the copper paddlewheel and the BTC linker, it could form a hydrophobic barrier that hinders water uptake (inside the otherwise hydrophilic pores). Due to this hydrophobic barrier, the activation time of the $I_2$@HKUST-1 crystal increases over other guests (under $N_2$ flow). The data clearly show that the presence of $I_2$ does not affect $\varepsilon'$ due to non-polar nature of molecular iodine.

Further to water molecules, we carried out measurements on other polar guest molecules, namely methanol and ethanol. We note that activated HKUST-1 crystals from all the measurements gave an averaged $\varepsilon' = 2.75 \pm 0.25$. Figures 3c-d show that for the first guest inclusion cycle, the $\varepsilon'$ for methanol and ethanol increased to 10.40 and 7.51, respectively. The second inclusion cycle resulted in a minor decline of their $\varepsilon'$ values to 9.92 and 7.35, respectively. Upon evacuation, however, their $\varepsilon'$ values only partially reverted to ~5.26 (cycle 1) and declined to 4.95 (cycle 2), instead of reverting to the initial value of ~2.75. This phenomenon can be explained by trapping of guest molecules in the pores of HKUST-1. Confined solvent molecules tend to form a weak hydrogen bonding with the framework[20] and with the adjacent molecules, therefore become harder to evacuate without heating. Because methanol molecules are more polar than ethanol,[24] the presence of high polar species as a guest in HKUST-1 pores will result in higher $\varepsilon'$, which we established in a systematic way at 1 MHz as: $H_2O$@HKUST-1 ($\varepsilon' \sim 60$) > methanol@HKUST-1 (~10) > ethanol@HKUST-1 (~7.5).

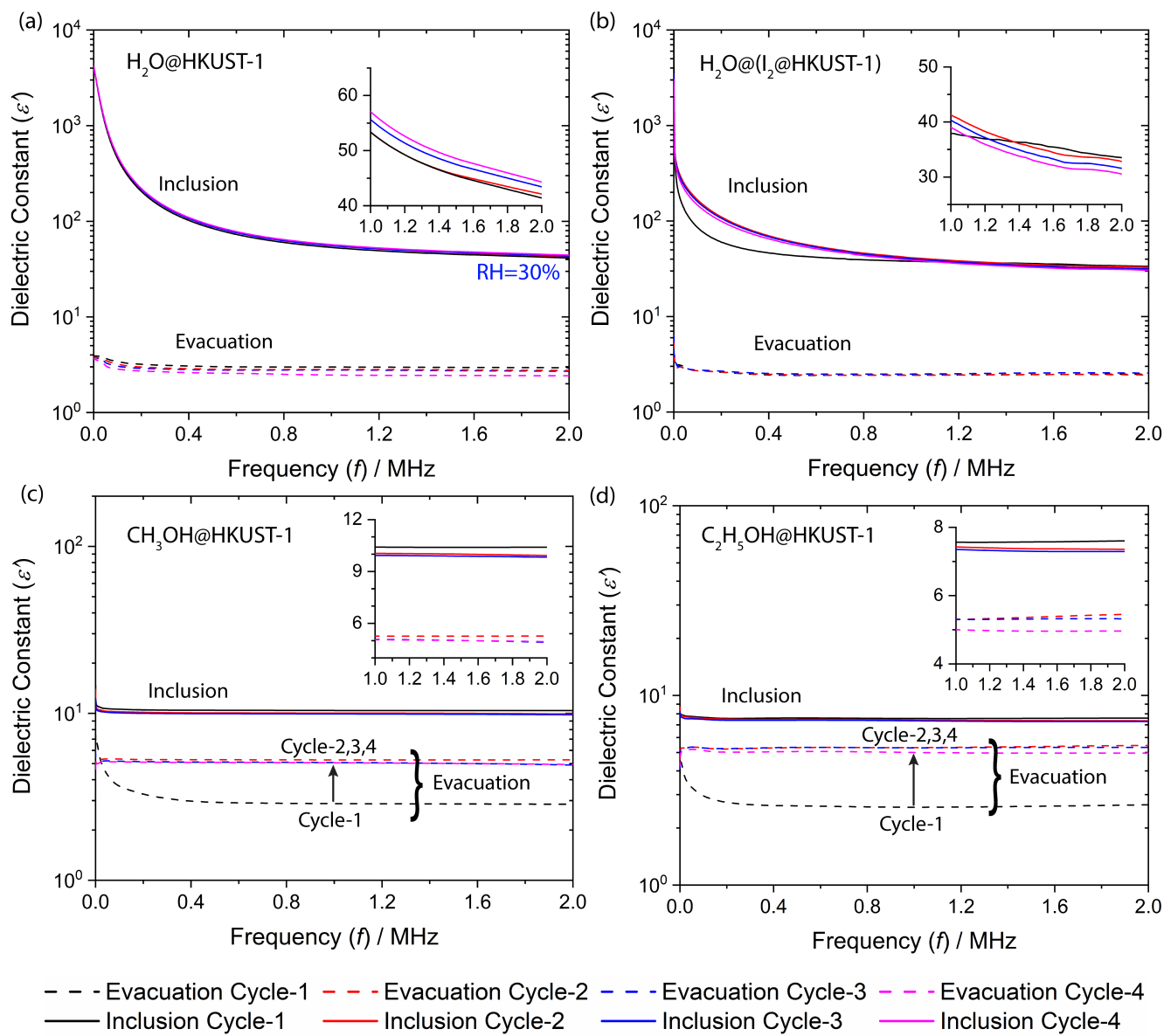

Figure 3: Effect of the encapsulation of different guest molecules into the HKUST-1 single crystal under a constant flow rate of 12 L/hr: (a) $H_2O$, (b) $H_2O$ and $I_2$, (c) methanol ($CH_3OH$) and (d) ethanol ($C_2H_5OH$), respectively, on the dielectric constant from 100 Hz to 2 MHz measured along the <100> lattice direction. The insets show the magnified views of the dielectric response in the range of 1-2 MHz.

Figure 4 shows the transient response of the dielectric and conductivity behavior of HKUST-1 crystal during guest inclusion-expulsion study at 1 MHz. The cyclic $\varepsilon'$ values in the encapsulated and activated states exceptionally match the $\varepsilon'$ values determined from the frequency sweep measurements of Figure 3. It can be seen that the cyclic data show good repeatability, when the induced cracks have stabilized after the second cycle (Figure S3). We found that encapsulation of non-polar $I_2$ affects the AC conductivity at 1 MHz, which increased from 0.16 mS m$^{-1}$ for activated HKUST-1 to 0.29 mS m$^{-1}$ for $I_2$@HKUST-1, and the latter increased to 7.9 mS m$^{-1}$ at 30% RH. Uptake of water molecules thus enhances AC conductivity of the $H_2O$@($I_2$@HKUST-1) crystal. HKUST-1 with methanol is relatively more conducting than ethanol, thus we determined 0.50 and 0.42 mS m$^{-1}$, respectively (Figures S9-S10).

Figure 5 is a representative plot of $\varepsilon'$ and its first-order derivative with time ($d\varepsilon'/dt$) for the first cycle, allowing a comparative study of rate-dependent response subject to different guests. The rise ($t_r$) and fall ($t_f$) times are summarized in Table 1. Unlike the inclusion rate, the expulsion rate of water from the crystal is significantly higher compared to the $I_2$-encapsulated framework. Similarly, the rise and fall times for the $I_2$-encapsulated framework is higher than the $I_2$-free crystal (Table 1); the data suggest that the hydrophobic barrier provided by molecular $I_2$ hinders the mobility of water molecules to and from the uncoordinated copper sites on the paddlewheel. The size of the guest molecule could also influence the inclusion-expulsion time and rate. Given the relatively larger size of ethanol, its mobility will be lower compared to other guest molecules considered, as reflected by magnitudes of $d\varepsilon'/dt$, ranked as: $C_2H_5OH < CH_3OH < I_2@H_2O < H_2O$ (Figure 5b inset).

*Table 1: Rise time and fall time* for inclusion and expulsion of guest molecules from a single crystal of HKUST-1 as the host framework.*

| Guest@HKUST-1 | Rise time, $t_r$ (sec) | Fall time, $t_f$ (sec) |
|---|---|---|
| $H_2O$ | 158.3 | 172.4 |
| $H_2O@I_2$ | 419.2 | 941.1 |
| $CH_3OH$ | 104.5 | 144.7 |
| $C_2H_5OH$ | 72.8 | 1088.6 |

*Defined as the time required for the signal to rise from its low value (10% of the step height) to high value (90% of the step height), and *vice versa*.

We reasoned that the larger guest molecules will fill the pores faster hence giving a lower rise time, while the clustering of spatially confined guests will impede their expulsion thus causing a higher fall time. Additionally, it is anticipated that weak chemical interaction between the guest species and the host framework will play a role in confinement of guest molecules (either strongly or weakly), which will determine the kinetics of cluster making-breaking process. This kinetics control the fall time of guest species being expelled from the pores. For instance, ethanol molecules can make a stronger hydrophobic interaction with the BTC linker, while interaction with methanol will be weaker comparatively. However, water will form strong bonding interaction with copper sites and due to absence of any hydrophobic groups, the interaction of water with the BTC linker will be relatively weaker compared with ethanol and methanol. On this basis, we could explain why the ethanol-encapsulated HKUST-1 exhibits a shorter rise time followed by a longer fall time over methanol (Table 1). Interestingly, our transient data also reveal that residual clusters of methanol and ethanol trapped in the pores (after the first inclusion cycle) will result in a lower inclusion rate in the subsequent cycles, see Figure S13(c)-(d).

In conclusion, we show that frequency-dependent dielectric measurement of a MOF single crystal is no longer insurmountable. Because of its high sensitivity for discriminating different molecular guests, this single-crystal dielectric approach will enable accurate characterization of transient phenomena. The ability to measure the dielectric response of an individual crystal yields intrinsic properties, which cannot be obtained by existing techniques relying on larger samples in the form of a compacted powder pellet or a polycrystalline sample. This is an important achievement because, to date, the dielectric characterization of pellets are affected by problems linked to unwanted framework deformation and/or phase amorphization from pelleting pressure. More generally, our results revealed that the dielectric response of MOF crystals can considerably be altered by host-guest interactions (at open metal sites or ligands) *via* hydrophobic-hydrophilic interaction, and by harnessing guest molecular size effect and guests clustering attributed to nanoconfinement. Finally, the ubiquitous dielectric properties and AC conductivity of MOFs bode well for future application, as a Guest@MOF sensor featuring vast tuneability and chemical selectivity, setting it ahead of the other competing resistive and fluorescent sensors in the field.

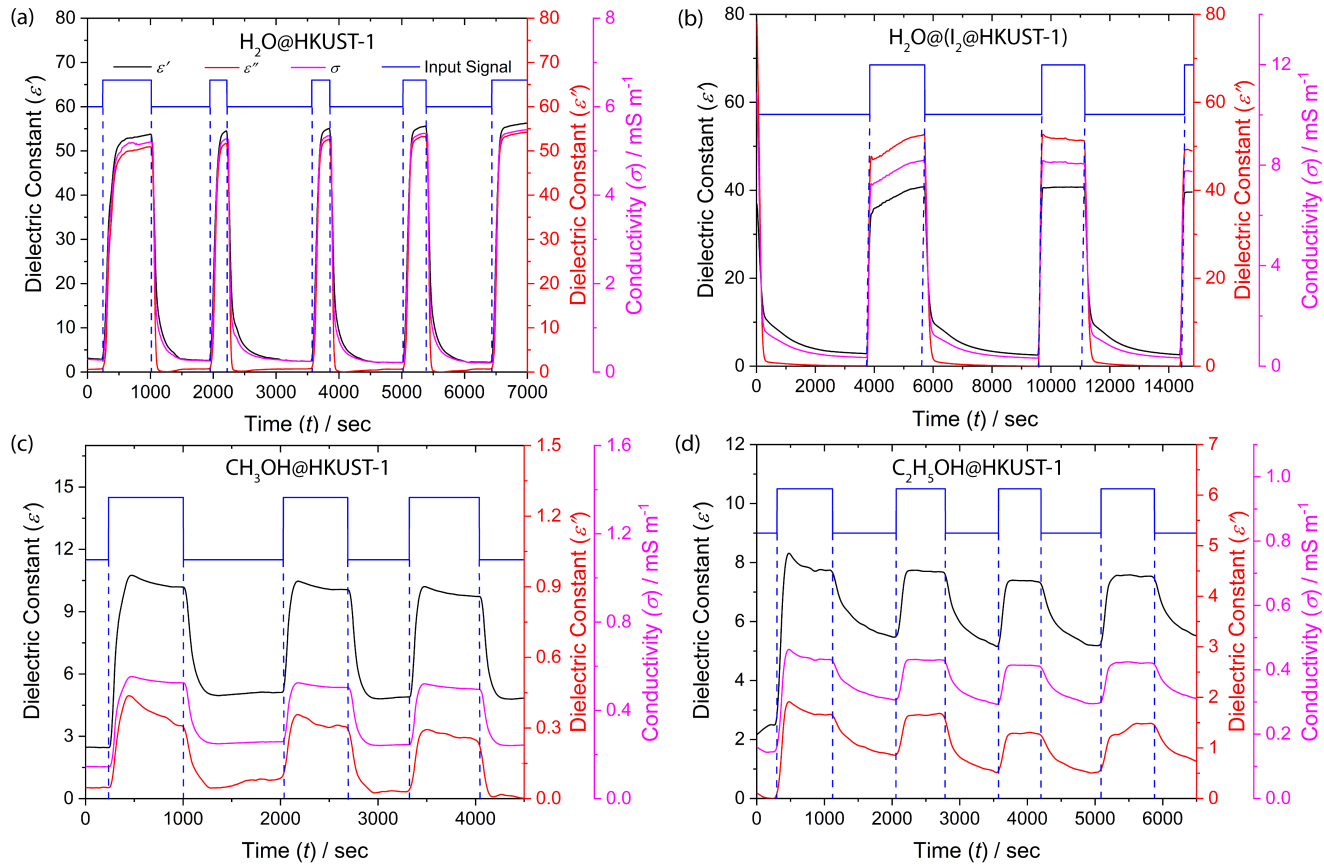

Figure 4: Cyclic dielectric (real and imaginary part) and conductivity measurements of HKUST-1 when being subjected to the following guest species (a) $H_2O$, (b) $H_2O$ and $I_2$, (c) methanol ($CH_3OH$) and (d) ethanol ($C_2H_5OH$). All measurements were performed at 1 MHz.

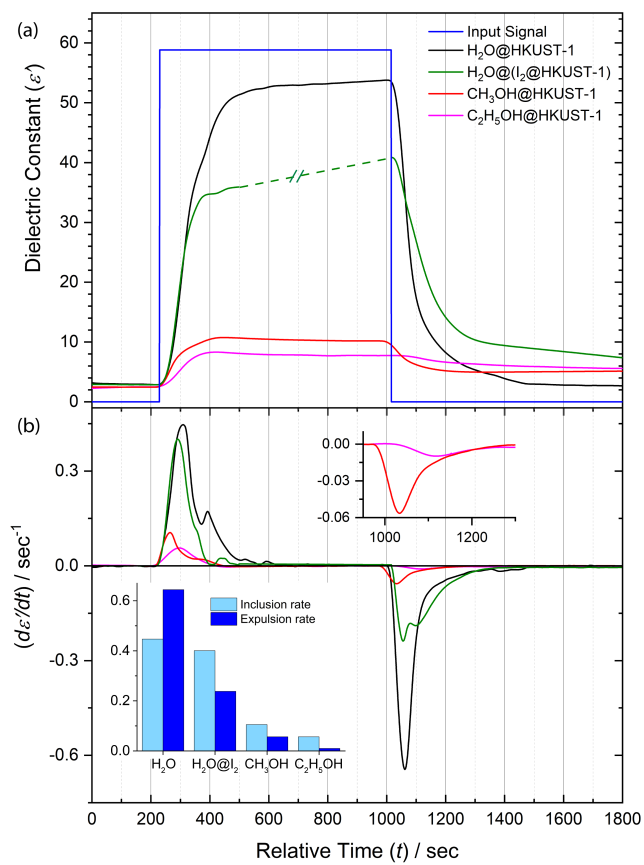

Figure 5: Comparison of time-dependent response for the inclusion and expulsion of different guest species in HKUST-1 single crystal. (a) First cycle of dielectric response under guest environment. (b) First-order derivative of the real part of dielectric constant with respect to time. The inset shows the inclusion and expulsion rates of different guest species. The $I_2$@HKUST-1 data were truncated on the time axis for better comparison with other guests.

## ASSOCIATED CONTENT

### Supporting Information

Experimental setup; Crystal coating and mounting procedures; Dielectric and AC conductivity of guest-encapsulated HKUST-1 crystal; Transient response guest inclusion-expulsion curves.

## ACKNOWLEDGMENTS

ASB is grateful to the Engineering Science (EPSRC DTP – Samsung) Studentship that supports this DPhil research. JCT and AKC acknowledge the European Union's Horizon 2020 research and innovation programme (ERC Consolidator Grant agreement No. 771575 - PROMOFS), the EPSRC Impact Acceleration Account Award (EP/R511742/1), and the Samsung GRO Award (DFR00230) for supporting this research. HHMY acknowledges the Glasstone Bequest for a Samuel and Violette Glasstone Fellowship. We acknowledge the Mechanical Workshop at Physical and Theoretical Chemistry laboratory, University of Oxford, for assistance in constructing the single-crystal cell. We thank Dr Mark Frogley and Dr Gianfelice Cinque at Diamond Light Source for the provision of Raman spectroscopy.